\documentclass[a4paper]{article}

\usepackage{amsthm,amsmath}
\usepackage{graphicx}
\graphicspath{ {images/} } \usepackage{mathtools}
\usepackage{mathrsfs}
\usepackage{epstopdf}
\usepackage[utf8]{inputenc} 

\let\OLDthebibliography\thebibliography
\renewcommand\thebibliography[1]{
  \OLDthebibliography{#1}
  \setlength{\parskip}{0pt}
  \setlength{\itemsep}{0pt plus 0.3ex}
}

\providecommand{\keywords}[1]{\textsc{Key words: } #1}

\graphicspath{ {images/} } \def\bSig\mathbf{\Sigma}
\newcommand{\bmath}[1]{\boldsymbol{#1}}

\title{Conditional Estimation in Two-stage Adaptive Designs}

\author
{Per Broberg \\
\small{Division of Cancer Epidemiology, Department of Clinical Sciences Lund, Lund University,}\\ \small{Skane University Hospital, Lund, Sweden. Email: per.broberg@med.lu.se.} 
\\[2mm]
Frank Miller \\
\small{Department of Statistics, Stockholm University, 10691 Stockholm, Sweden.}\\ \small{Email: frank.miller@stat.su.se.}}

\begin{document}

\date{ }

\maketitle

\begin{abstract}
We consider conditional estimation in two-stage sample size adjustable designs and the consequent bias.
More specifically, we consider a design which permits raising the sample size when interim results
look rather promising, and which retains the originally planned sample size when results look very promising.
The estimation procedures reported comprise the unconditional maximum likelihood, the conditionally unbiased Rao-Blackwell estimator, the conditional median unbiased estimator, and the conditional maximum likelihood with and without bias correction.
We compare these estimators based on analytical results and a simulation study. 
We show how they can be applied in a real clinical trial setting.
\end{abstract}

\\[15mm]

\noindent
This is the peer reviewed version of the following article: ``Conditional Estimation in Two-stage Adaptive Designs'', which has been published in final form at Biometrics with DOI 10.1111/biom.12642. Link: {\tt \bf http://onlinelibrary.wiley.com/doi/10.1111/biom.12642/full}. This article may be used for non-commercial purposes in accordance with Wiley Terms and Conditions for Self-Archiving.\\[15mm]

\noindent
Reference for the published article at Biometrics:\\
Broberg P, Miller F (2017).
Conditional estimation in two-stage adaptive designs. {\it Biometrics}, {\bf 73}, 895-904.
\newpage

\section{Introduction}
\label{s:intro}

Adaptive design has made its way into the repertoire of practical clinical biostatistics as shown for example by the review of \cite{Bretz2009} and the discussion by \cite{Iasonos2014}. European and US guidance documents for the use of adaptive designs in clinical trials have been developed.
Adaptive designs are applied in the confirmatory context (Phase III) as well as in earlier development. For example, adaptive Phase II or seamless Phase II/III case studies have been described by \cite{Miller2014189} and \cite{Cuffe2014}.

In this article, we consider recalculation of the sample size based on an interim effect estimate. 
For this adaptive design, statistically valid methods are required for inference. Much work has been done in recent decades to provide significance tests which adequately control the type I error rate in the context of sample size recalculation. For example, conditions have been identified by \cite{Posch2003} and 
\cite{broberg2013} where the conventional significance test (which would be done in a non-adaptive situation) still controls the type I error rate after sample size recalculation. 
In these references, the general conditional error approach is applied, which states that adaptations are allowed as long as the conditional error rate of the final test is not increased by the adaptation, see \cite{Mueller2001, Mueller2004}. Alternatively, if it is not desired to adhere to these conditions, several good methods exist to control the type I error with modified significance 
tests in this adaptive setting.

However, point and interval estimates are usually of high importance for clinical interpretation of trial results. It has been identified that inference in 
adaptive designs needs to focus more on estimation, which according to \cite{Bretz2009} is a topic of ongoing research. 



Estimation has been considered for treatment selection designs where, in an interim analysis, treatment(s) with best (largest) estimates are selected for Stage 2. It is well known that the naive effect estimate of the treatment with best interim result is positively biased due to selection. This and the bias after more complex selection situations is discussed by  \cite{bauer2010}. To tackle this problem, estimation under the condition of the order statistics of interim means has been investigated. \cite{Cohen1989273} derived a conditional Rao-Blackwell estimator. \cite{SIM:SIM5757} generalised this to allow for a futility stop and developed an estimator correcting the unconditional maximum likelihood estimate (MLE) for its conditional bias;  \cite{robertson2016} generalized to a multivariate normal setting with known covariance structure. \cite{carreras2013} and \cite{bowden2014} use shrinkage estimators to address selection bias.

In contrast to the mentioned publications on treatment selection, we investigate here an interim decision rule for sample size recalculation where sample size depends on the estimated effect in Stage 1. Instead of treatment selection bias, a bias can occur here in the unconditional case due to data dependent weighting of the stages. In this article we focus on inference under the condition that a specific sample size is chosen. \cite{SIM:SIM4430} consider the case of two possible sample sizes: futility stop or continue to a prespecified sample size. They present conditional Rao-Blackwell, conditional median unbiased and conditional MLE. We extend the design allowing for a finite (usually low) number of possible sample size decisions. In the \cite{SIM:SIM4430} setting it is heuristically clear that the three conditional estimators are smaller than the unconditional MLE under the condition of continuation. In contrast, the relation between the estimates is not obvious for the sample size recalculation design of this paper. We will prove a result pinpointing when the conditional estimators are larger or smaller than the unconditional MLE leading to a general insight how these conditional estimates behave. 

Further to the estimators considered by \cite{SIM:SIM4430} we motivate and investigate here a new estimator where we bias-correct the conditional MLE,  employing a new bias approximation using the third derivative of the log-likelihood.

In contrast to the mentioned publications on treatment selection, we will compare here all four  conditional estimators mentioned above and the unconditional MLE head-to-head with regard to their bias, variance and mean squared error (MSE) by simulations for our sample size recalculation design. Unlike these treatment selection publications and unlike \cite{SIM:SIM4430}, we compare not only average measures (bias, variance, MSE) but also differences between the estimators on an individual data level as we describe the conditional estimators as functions of the unconditional MLE. We can order conditional Rao-Blackwell, conditional median unbiased and conditional MLE in terms of how much they differ from MLE.



In this article, we will consider 
the case of 
normally distributed data with known variance. In Section \ref{ch2}, we provide the general assumptions and the considered sample size recalculation rule, discuss bias of the MLE and mention unconditionally unbiased estimates. We derive several conditional estimators in Section \ref{ch3}  and compare them 
in Section \ref{ch4}. Section \ref{sec_extensions} provides extensions to our model. For a previously conducted randomized controlled clinical trial which investigated a treatment for schizophrenia patients, we show how the considered estimators can be applied
in Section \ref{ch_ex}. 
A discussion concludes in Section \ref{ch5}.


\section{Background}
\label{ch2}

\subsection{General setting and bias}

We consider the one-sample case with independent normally distributed observations.
Let the observations be $X_1, X_2, \dots \sim N(\mu,\sigma^2)$ with $\sigma^2$ known. 
An interim analysis is performed after $n_1$ observations (Stage 1) and based on the results, the sample size $N_2$ for Stage 2 is determined; the total sample size is $N=n_1+N_2$ (note that $N_2$ and $N$ are random variables). We use the simplified notation $Y_{1}=\sum_{i=1}^{n_1}X_{i}/n_{1}$, $\sigma_{1}=\sigma/\sqrt{n_{1}}$; $Y_{2}=\sum_{i=n_1+1}^{N}X_{i}/N_{2}$ represents the additional information collected after the interim analysis, $\sigma_{2}=\sigma/\sqrt{N_{2}}$. 


In a study with the aim to test $H_0: \mu \leq 0$ versus $H_1: \mu>0$, the parameter $\mu$ should be estimated by a $\widehat \mu$ after all $N$ observations.
The mean of all observations is the unconditional maximum likelihood estimate (MLE) for $\mu$. Therefore, we write $\widehat{\mu}_{ML}=Y=\sum_{i=1}^{N}X_{i}/N$ and further, $\sigma_0=\sigma/\sqrt{N}$. Note that $\sigma_0, \sigma_1, \sigma_2$ (as well as $\sigma_A$ and $\sigma_B$ introduced later) are all multiples of $\sigma$ which we use to improve readability of formulae.

\cite{SIM:SIM2258} show that estimators of the form $\widehat\mu_{WM,w}=w(Y_{1})Y_{1}+\{1-w(Y_{1})\}Y_{2}$ with $w(Y_1)\in [0,1]$, i.e. a weighted mean of $ Y_{1}$ and $Y_{2}$ where the weight $w$ may depend on the first stage outcome, have bias $Cov\left\{w(Y_{1}), Y_{1}\right\}$ for all $\mu$.

Choosing the Stage 2 sample size $N_2=N_2(Y_1)$ as a decreasing function of $Y_1>0$ is a natural choice as larger sample sizes are needed if the true $\mu>0$ is small. If one uses $\widehat\mu_{ML}$ which is equal to $\widehat \mu_{WM,w}$ with $w(Y_1)=n_1/\{n_1+N_2(Y_1)\}$, the result of \cite{SIM:SIM2258} therefore shows that $\widehat \mu_{ML}$ overestimates $\mu$ under $\mu>0$.

In this paper, we consider the following design: We decide in the interim analysis to stop for futility if the observed effect $Y_1$ is low (decision $R=0$ and final sample size $n_{tot}^{(0)}\geq n_1$); otherwise we make one out of $s$ decisions where decision $R=t$ means that sample size $n_{tot}^{(t)}, t=1,\dots,s,$ is chosen depending on in which of $s$ predefined intervals $Y_1$ is. Formally,
\begin{equation}
\label{two}
  R = t \mbox{ and }
  N = N(Y_1) = n_{tot}^{(t)} 
  \mbox{ if } c_t < Y_1 \leq c_{t+1}, \quad t=0,1,\dots,s 
\end{equation}
for $s\geq 2$ and some constants $c_0=-\infty \leq c_1 < \dots < c_s \leq c_{s+1} = \infty$.

To simplify the presentation, we subsequently consider the case $s = 2$ -- however, all  our results are directly applicable for $s>2$ as well. We write therefore:
\begin{equation}\label{three}
  R = \left\{ 
 \renewcommand{\arraystretch}{0.8}
 \begin{array}{l}
    0 \\
    1 \\
    2 
  \end{array} \right.
  \mbox{ and }
  N = N(Y_1)= \left\{ 
  \renewcommand{\arraystretch}{0.8}
  \begin{array}{l l}
    n_{f} & \quad \textrm{if } Y_{1} \le c_{1} \\
    n_{\max} & \quad \textrm{if } c_{1} < Y_{1} \le c_{2}\\
    n_{0} & \quad \textrm{if } Y_{1} > c_{2}
  \end{array} \right.
\end{equation}
where $n_0=n_{tot}^{(2)}$, the originally planned sample size, $n_{\max}=n_{tot}^{(1)}$, the maximal sample size increased from the planned (since observed effect was promising but limited), and $n_f=n_{tot}^{(0)}$, the sample size in case of a futility stop. Usually $n_f=n_1$ is desired but sometimes this is not possible in practice and an ``overrun'' $n_f>n_1$ needs to be accepted.
Note that a common two-stage group sequential design without sample size recalculation is an important special case of (\ref{three}) with $n_f=n_0=n_1$ (or allowing overrun, $n_f=n_0\geq n_1$).

In the design (\ref{three}), the bias of $\widehat \mu_{ML}=\widehat \mu_{WM,w}$ with $w(Y_1)=n_1/N=n_1/N(Y_1)$ is:
\begin{eqnarray*}
Cov\left\{w(Y_{1}), Y_{1}\right\} 
& = & E\left\{w(Y_{1}) Y_{1} \right\}-E\left\{w(Y_{1})\right\} E\left( Y_{1} \right)
=n_{1}E\left(Y_{1}/N\right)-E\left(n_{1}/N\right)E\left(Y_{1}\right) \\
&=& n_{1}\sigma_{1}\left\{-\frac{\phi(\frac{c_{1}-\mu}{\sigma_{1}})}{n_{1}}-\frac{\phi(\frac{c_{2}-\mu}{\sigma_{1}})-\phi(\frac{c_{1}-\mu}{\sigma_{1}})}{n_{\max}}+\frac{\phi(\frac{c_{2}-\mu}{\sigma_{1}})}{n_{0}}\right\},
\end{eqnarray*}

where $\phi$ and $\Phi$ denote the density and the distribution function of the standard normal distribution, respectively.

\begin{figure}
\centering
\includegraphics[width=160mm]{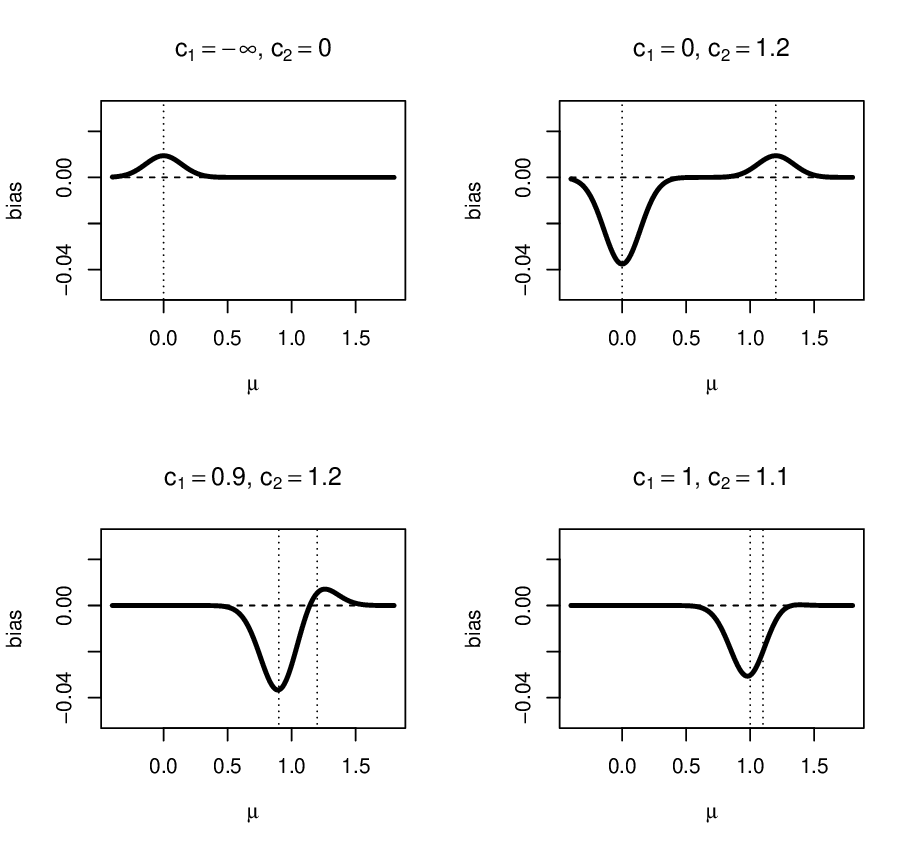}
\caption{Bias of the maximum likelihood estimator $\widehat \mu_{ML}$ in dependence of $\mu$ for different values of $c_1, c_2$ as indicated above each panel and for $n_1=50, n_0=100, n_{\max}=150,\sigma=1$. The vertical dotted lines indicate $c_1$ and $c_2$.}
\label{fig_bias_uncond}
\end{figure}

For example for $n_1=50, n_0=100, n_{\max}=150, \sigma=1$, the bias of $\widehat \mu_{ML}$ in dependence of $\mu$ is shown in Figure \ref{fig_bias_uncond} for some choices of $(c_1, c_2)$. The bias is negative around $c_1$ (where the sample size increases from $Y_1<c_1$ to $Y_1>c_1$) and is positive around $c_2$ (where the sample size decreases from $Y_1<c_2$ to $Y_1>c_2$). When $c_1$ and $c_2$ are very close (e.g. $c_1=1, c_2=1.1$, lower right panel), the positive bias can almost be hidden by the overlying negative bias. 

\subsection{Unconditionally unbiased estimates of the treatment effect}

We can search for estimates $\widehat \mu$ which are unbiased in the traditional, unconditional sense, i.e. $E(\widehat \mu)=\mu$, by choosing appropriate weight functions $w(Y_1)$ for $\widehat \mu_{WM,w}$.
If the trial cannot stop at the interim, i.e. $c_1=-\infty, n_{max}>n_1, n_0 > n_1$, then any predetermined weights, $w(Y_1)=w_{1}$ and $1-w_1$,
will ensure an unbiased estimate
$
\widehat{\mu}=w_{1}Y_{1}+(1-w_{1})Y_{2}.
$
According to \cite{Chang2007}, a good choice are weights
$w_{1}=2n_{1}/(n_{0}+n_{max})$.
For designs permitting early stopping,  \cite{lawrencehung2003} present the median unbiased estimate 
$
\widehat{\mu}=(w_{1}\sqrt{n_{1}}Y_{1}+w_{2}\sqrt{N_{2}}Y_{2})/(w_{1}\sqrt{n_{1}}+w_{2}\sqrt{N_{2}}),
$
for predetermined weights $w_{1}, w_{2}$, suggested to be  $w_{1}=\sqrt{n_{1}/n_{0}}$ and $w_{2}=\sqrt{1-w_{1}^{2}}$.
We will subsequently not restrict ourselves to predetermined weights, but will permit the weights to change due to adaptations.

\section{Conditional estimation after sample size recalculation}
\label{ch3}

If the consequences of bias are different depending on the interim decision $R$, conditional estimation is sensible.
In studies with an interim decision about futility stop only ($c_2=\infty$, i.e. $R\in\{0,1\}$), it is reasonable to require unbiasedness specifically if the trial is continued to Stage 2 ($R=1$). Only if not stopped for futility, results are used by regulatory agencies for decisions about licensing or if a Phase II trial is considered by the sponsor for decisions about continuation of the program to Phase III. Therefore, \cite{pepe2009} advocate requiring unbiasedness under the condition that the trial is continued, $E(\widehat \mu|R=1)=\mu$. 

However, in our situation with decision rule (\ref{three}), i.e. when based on an observed large effect the study is continued to a smaller sample size or stopped directly, the situation is a little more complex. The interesting situations when good properties of estimators are required are both $R=1$ and $R=2$. We therefore investigate the bias under the condition of $R$ and show how to make the analysis unbiased in the conditional setting.
Cases $R=1$ and $R=2$ are of main interest, hence we do not subsequently consider  $R=0$.

\subsection{The unconditional MLE}
\label{sec_mle}

We will first consider the unconditional MLE $\widehat{\mu}_{ML}$ ($=Y$). The conditional distribution is presented in Section \ref{sec_cmu}.
The conditional bias of $\widehat{\mu}_{ML}$ equals
\begin{equation}\label{eq:biasml}
  bias_{ML}(\mu, R) = \left\{ 
  \begin{array}{ll}
  \sigma\frac{\sqrt{n_{1}}}{N}\frac{\phi(\overline{b}_{1})-\phi(\overline{b}_{2})}{\Phi(\overline{b}_{1})-\Phi(\overline{b}_{2})}, & R=1,\\
  \sigma\frac{\sqrt{n_{1}}}{N}\frac{\phi(\overline{b}_{2})}{\Phi(\overline{b}_{2})}, & R=2, 
  \end{array}
  \right.
\end{equation}
where $\overline{b}_i=\overline{b}_{i}(\mu)=(\mu-c_{i})/\sigma_1, i=1,2$, cf. (13.133) and (13.134) in \cite{jkb1994}. 
When $R=1$ we have the following relations. If $\mu < (c_{1}+c_{2})/2$, then $bias_{ML} > 0$. If $\mu > (c_{1}+c_{2})/2$, then $bias_{ML} < 0$. Whenever $R=2$ there is a positive bias.
However, since $\mu$ is unknown we cannot calculate this bias in practice. There is another way to construct an unbiased estimator which builds on the Rao-Blackwell theorem.

\subsection{Uniformly Minimum Variance Unbiased Estimation}
\label{sec32}

In the context of treatment selection designs, the Rao-Blackwell theorem has been used to find an
Uniformly Minimum Variance Unbiased Estimator, under the condition of the order statistics of interim means, \cite{Cohen1989273, BIMJ:BIMJ200810442}. \cite{SIM:SIM5757} have additionally conditioned on proceeding to Stage 2 since they allowed for futility stopping. For further generalisations see \cite{bowden2014a, robertson2016}.
In these treatment selection designs, the sample size of Stage 2 is given and one proceeds to Stage 2, selecting the best treatment(s) and dropping the others.   

We will now derive results for the case of our sample size recalculation design without selection. 
We condition here on the interval which the observed interim effect is in and not on an order statistic of interim means.

The following theorem presents the Uniformly Minimum Variance Conditionally Unbiased Estimator (UMVCUE) and its conditional variance.
The estimator consists of two terms: the unconditional MLE $\widehat{\mu}_{ML}$ and a product of a standard deviation and a ratio involving $\phi$ and $\Phi$, which in the case of $R=2$ is equal to the inverse Mills ratio $\nu(x)=\phi(x)/\Phi(x)$. The unbiasedness is achieved by subtracting a term which has a structure similar to the bias (\ref{eq:biasml}). The proof can be found in Web Appendix A.

{\bf Theorem:}
The UMVCUE $\widehat \mu_{RB}$ is given by
\begin{equation}
\label{eq_rb}
  \widehat \mu_{RB} = \left\{ 
  \begin{array}{ll}
  g_{1}(\widehat{\mu}_{ML})=  \widehat{\mu}_{ML}-\sigma_{B}\frac{\Delta \phi}{\Delta \Phi}, & R=1,\\
g_{2}(\widehat{\mu}_{ML})= \widehat{\mu}_{ML} - \sigma_B \nu(z_2), & R=2, 
  \end{array}
  \right.
\end{equation} 
where $\Delta \phi=\phi(z_1)-\phi(z_2)$, $\Delta \Phi = \Phi(z_1)-\Phi(z_2)$, $z_i=z_i(\widehat{\mu}_{ML})=(\widehat{\mu}_{ML}-c_{i})/\sigma_A, i=1,2$, and $\sigma_{A}^{2}=\sigma_{1}^{4}/(\sigma_{1}^{2}+\sigma_{2}^{2}), \sigma_{B}^{2}=\sigma_{2}^{4}/(\sigma_{1}^{2}+\sigma_{2}^{2})$.
The conditional variance is given by:
\begin{equation}
\label{eq_varrb}
  Var(\widehat \mu_{RB}|R) \approx \left\{ 
  \begin{array}{ll}
    g_{1}'(\mu_{1})^{2}Var(\widehat{\mu}_{ML}|R=1), & R=1, \\
   g_{2}'(\mu_{2})^{2}Var(\widehat{\mu}_{ML}|R=2), & R=2,
  \end{array}
  \right.
\end{equation}
where 
\begin{eqnarray*}
  g_{1}'(y) & = & 1+\frac{\sigma_{2}^{2}}{\sigma_{1}^{2}}\frac{\left\{z_1(y)\phi(z_1(y))+z_2(y)\phi(z_2(y))\right\}\Delta \Phi + \Delta \phi^{2}}{\Delta \Phi ^{2}}, \\
  g_{2}'(y) & = & 1+\frac{\sigma_{2}^{2}}{\sigma_{1}^{2}}\frac{z_2(y)\phi(z_2(y))\Phi(z_2(y))+\phi(z_2(y))^{2}}{\Phi(z_{2}(y)) ^{2}},
\end{eqnarray*}
and the variance factors in (\ref{eq_varrb}) are detailed in the proof in formulae (A.1) and (A.2) in Web Appendix A. 

Note that the values of $\sigma_2, \sigma_A, \sigma_B$ are different for case $R=1$ and $R=2$ since they depend on $N_2$.

\begin{figure}[htb]
\centering
\includegraphics[width=120mm]{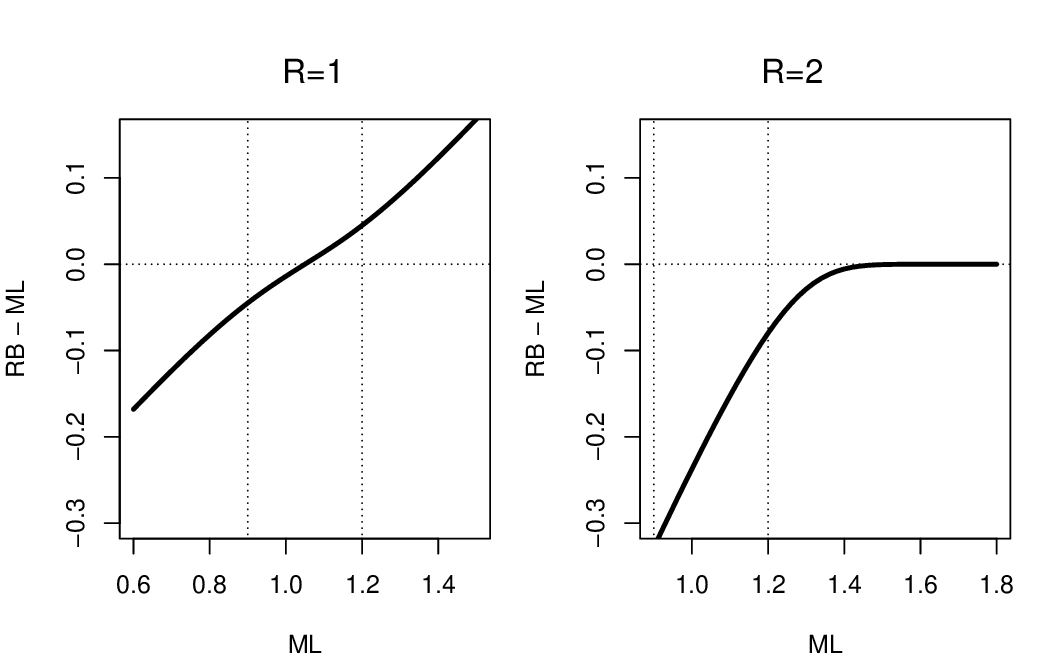}
\caption{Difference between Rao-Blackwell (RB) and maximum likelihood (ML) estimator versus ML estimator for $R=1$ (left) and $R=2$ (right). Here $n_{1}=50, n_{0}=100, n_{max}=150, c_{1}=0.9, c_{2}=1.2, \sigma=1$. The vertical dotted lines indicate $c_1$ and $c_2$.}
\label{fig_rb_est}
\end{figure}

To illustrate the difference between the Rao-Blackwell estimate in the above theorem and the MLE, we show $\widehat \mu_{RB}-\widehat \mu_{ML}$ versus $\widehat \mu_{ML}$ for $R=1$ and $R=2$, see Figure \ref{fig_rb_est}, with the specific values $n_1=50, n_0=100, n_{max}=150, c_1=0.9, c_2=1.2$ and $\sigma=1$. For example let us look at $R=1$, i.e. when we have $0.9<Y_1<1.2$ for the interim mean. Then the total sample size is raised from $n_0=100$ to $n_{max}=150$. If the final mean, $\widehat \mu_{ML}$, is still between 0.9 and 1.2, the MLE and the Rao-Blackwell differ by at most 0.05. If $\widehat \mu_{ML}>1.2$, then the Rao-Blackwell estimate is larger than the MLE and if $\widehat \mu_{ML}<0.9$, then the Rao-Blackwell estimate is smaller than the MLE; this means in both cases: $\widehat \mu_{RB}$ is shrunken towards $Y_2$.


In Figure \ref{fig_rb_est} for $R=1$, $\widehat \mu_{RB}-\widehat \mu_{ML}>0$ if and only if $\widehat \mu_{ML}>1.05$. Note that 1.05 is the mean of $c_1$ and $c_2$. Further we observe that $\widehat \mu_{RB}-\widehat \mu_{ML}<0$ for $R=2$. These observations are valid in general, see Section \ref{sec_comp}.

Let us consider in general what happens for the case $R=1$ if the interval $(c_1,c_2]$ becomes small. If $c_2\to c_1$, $\widehat \mu_{RB}$ will approach $Y_2$. This follows as $\widehat{\mu}_{RB}=E(Y_{2}| Y, Y_{1} \in I )=E(Y_{2}| Y, Y_{2} \in J )$, see Web Appendix A.

\subsection{Conditional median unbiased Estimation}
\label{sec_cmu}

The conditional median unbiased estimation comes naturally from interval estimation and is the solution with respect to $\mu$ for $q=0.5$ of the equation 
\begin{equation} \label{cmu_def}
\int_{-\infty}^{\widehat\mu_{ML}}f\left(y|\mu, R \right)dy=q,
\end{equation}
cf. \cite{SIM:SIM4430} and \cite{Whitehead1992}. It will be denoted by $\widehat\mu_{CMU}$. Note that interval estimation can be achieved by solving for $q_{1}$ and $q_{2}$  with $q_{2} > 0.5 > q_{1}$ such that $q_{2}-q_{1}=l$ is the desired confidence level.

Take the case $R=1$ first. As in the derivation of the Rao-Blackwell estimator we may show that the joint density for $Y_1$ and $Y$ under the condition may be written as
$$
f(y_1,y)=\widetilde k(\mu) \phi\left(\frac{y-\mu}{\sigma_0}\right)\phi\left(\frac{y_{1}-y}{\sigma_A}\right)
{\bf 1 }\{c_1 < y_{1} \leq c_2\}
$$
with a function $\widetilde k(\mu)$ independent of $y_1$ and $y$. Integrating out $y_{1}$ yields the conditional density: 
\[ 
f(y|\mu, R=1)=\frac{\Phi\left(\frac{c_{2}-y}{\sigma_{A}}\right)-\Phi\left(\frac{c_{1}-y}{\sigma_{A}}\right)}{\Phi\left(\frac{c_{2}-\mu}{\sigma_{1}}\right)-\Phi\left(\frac{c_{1}-\mu}{\sigma_{1}}\right)}
\cdot \frac{\phi\left(\frac{y-\mu}{\sigma_{0}}\right)}{\sigma_{0}},
\]   
see also (13.133) in \cite{jkb1994}.
From this the case $R=2$ follows:
\[ 
f(y|\mu, R=2)=\frac{1-\Phi\left(\frac{c_{2}-y}{\sigma_{A}}\right)}{1-\Phi\left(\frac{c_{2}-\mu}{\sigma_{1}}\right)}
\cdot \frac{\phi\left(\frac{y-\mu}{\sigma_{0}}\right)}{\sigma_{0}},
\]   
cf.\ equation (2) in \cite{SIM:SIM4430}. 

Building on \cite{SIM:SIM4430} one may derive an approximate bias formula for $\widehat{\mu}_{CMU}$. 
Using equation (8) in \cite{SIM:SIM4430} we write
$$
0.5\approx F(\widehat{\mu}_{ML}|\mu, R)+(\widehat{\mu}_{CMU}-\mu)\frac{\partial F(\widehat{\mu}_{ML}|\mu, R)}{\partial \mu}.
$$
To denote the conditional expectation of $\widehat{\mu}_{ML}$ use the notation $\mu_{R}=\mu+bias_{ML}(\mu, R)$, where the bias is specified in equation (\ref{eq:biasml}).
This means that the bias approximately equals 
\begin{equation}\label{eq:cmubias}
bias_{CMU}(\mu, R)=E(\widehat{\mu}_{CMU}-\mu|\mu,R) \approx \frac{0.5- F(\mu_{R}|\mu, R)}{\frac{\partial F(\mu_{R}|\mu, R)}{\partial \mu}}.
\end{equation}
For $R=1$, the derivative equals
\begin{equation}\label{eq:derFmu}
\frac{\partial F(x|\mu, R=1)}{\partial \mu}=\int_{-\infty}^{x}f(t|\mu, R=1) \left[\frac{t-\mu}{\sigma_{0}^{2}}+\frac{\phi(\frac{c_{2}-\mu}{\sigma_{1}})-\phi(\frac{c_{1}-\mu}{\sigma_{1}})}{\sigma_{1}\left\{\Phi(\frac{c_{2}-\mu}{\sigma_{1}})-\Phi(\frac{c_{1}-\mu}{\sigma_{1}})\right\}}\right]dt.
\end{equation}
Then evaluate at  $x=\mu_{1}$, with $\mu_{1}=\mu+bias_{ML}(\mu, 1)$, and similarly for $R=2$. 
%
The delta method provides an estimate of the variance through
$$
Var(\widehat{\mu}_{CMU}|\mu, R)\approx \left\{\frac{\partial}{\partial x}\frac{0.5-F(x|\mu, R)}{\frac{\partial F(x|\mu, R)}{\partial \mu}}|_{x=\mu_{R}}\right\}^{2}Var(\widehat{\mu}_{ML}|\mu, R).
$$
%
%

\subsection{Conditional maximum likelihood estimation}
\label{cMLE}

\cite{bebu2010, bebu2013} address conditional maximum likelihood estimation in a two-stage adaptive design with treatment selection. Here, instead, we permit two kinds of continuation to the second stage, either with or without sample size adjustment, and, we derive an explicit bias correction formula.  

To derive the likelihood function given $R=1$, we write $b_i=b_i(\mu)=\sqrt{n_{1}}(c_i-\mu)/\sigma, i=1,2,$ and then,
\begin{equation}\label{phid}
P(R=1)=P(c_1<Y_1\leq c_2)=\Phi(b_2(\mu))-\Phi(b_1(\mu)).
\end{equation}
The likelihood function is the product of a constrained multivariate distribution (for Stage 1) and an unconstrained (for Stage 2):
$$
L_R(\mu)=
\left\{\prod_{j=1}^{n_{1}}\frac{1}{\sigma}\phi\left(\frac{x_{j}-\mu}{\sigma}\right)\right\} \cdot \frac{{\bf 1}\{R=1\}}{P(R=1)} \times
\prod_{j=n_{1}+1}^{N}\frac{1}{\sigma}\phi\left(\frac{x_{j}-\mu}{\sigma}\right).
$$
Then the log likelihood becomes
$$
\mathscr{L}_R(\mu)=-N\log(\sqrt{2\pi}\sigma) -\sum_{j=1}^{N}\frac{(x_{j}-\mu)^{2}}{2\sigma^{2}}-\log(\Phi(b_2(\mu))-\Phi(b_1(\mu))).
$$
Taking the derivative $\mathscr{L}'_R(\mu)=\partial\mathscr{L}_R(\mu)/\partial \mu$ with respect to $\mu$ and using $b_i'(\mu)=-\sqrt{n_{1}}/\sigma$ yields
\begin{equation}\label{eq:mle1}
\mathscr{L}'_R(\mu)=\sum_{j=1}^{N}\frac{x_{j}-\mu}{\sigma^{2}}
+\frac{\sqrt{n_1}}{\sigma}\frac{\phi\left(b_{2}(\mu)\right)-\phi\left(b_{1}(\mu)\right)}{\Phi\left(b_{2}(\mu)\right)-\Phi\left(b_{1}(\mu)\right)}.
\end{equation}
Note that the last term of $\mathscr{L}'_R(\mu)$ equals $-N\times bias_{ML}(\mu, R)/\sigma^{2}$, and that the expected value of the first is  $N \times bias_{ML}(\mu, R) / \sigma^2$. So, $E(\mathscr{L}'_R(\mu)|R)=0$, cf. Lemma 6.1 of \cite{lehmann1983}.

Solving the optimality condition
\begin{equation}
\label{eq_cml}
\sigma^{2}\mathscr{L}'_R(\mu)=N(\widehat\mu_{ML}-\mu)+\sqrt{n_{1}}\sigma\frac{\phi\left(b_2(\mu)\right)-\phi\left(b_1(\mu)\right)}{\Phi\left(b_2(\mu)\right)-\Phi\left(b_1(\mu)\right)}=0
\end{equation}
with respect to $\mu$ yields the conditional maximum likelihood estimate $\widehat{\mu}_{CML}$. 

An estimate of the variance of the estimator may be obtained through the observed information $j$
$$
j(\widehat{\mu}_{CML})=\mathscr{L}''_R(\mu)|_{\mu=\widehat{\mu}_{CML}}
= -\frac{N}{\sigma^{2}}
+\frac{b_2 v_2 - b_1 v_1}{s r} 
+ \frac{(v_2-v_1)^{2}}{r^2}|_{\mu=\widehat{\mu}_{CML}},
$$
with $s=\sigma/\sqrt{n_{1}}$, $v_i=\phi\left(b_i(\mu)\right)/s$, $b_i=b_i(\mu)$ and $r=P(R=1)=\Phi(b_2)-\Phi(b_1)$. 
This gives us the  variance estimate  $-j(\widehat{\mu}_{CML})^{-1}$. Since $\widehat{\mu}_{CML}$ asymptotically attains the the Cram\'{e}r-Rao bound and unbiasedness, one could consider estimating also $Var(\widehat{\mu}_{RB}|R)$ by $-j(\widehat{\mu}_{CML})^{-1}$. For a comparison of different estimators see \cite{lane2014}.

%

As explained in \cite{cox1994} the bias of the MLE is related to the first three derivatives of the log likelihood. The derivation in Web Appendix A gives the following bias formula
(for a closed form presentation of $\mathscr{L}'''_R(\mu)$ see Web Appendix A): \begin{equation}
\label{eq:biasdef}
bias_{CML}(\mu,R)= \mathscr{L}'''_R(\mu) / \{2\mathscr{L}''_R(\mu)^2\} +\mathcal{O}(N^{-3/2}).
\end{equation}

A bias corrected estimate $\widehat\mu_{CMLc}$ comes from solving the equation  
\begin{equation}\label{eq:bias}
\widehat \mu_{CML}=\mu+bias_{CML}(\mu,R)
\end{equation}
with respect to $\mu$, cf. equation (5.6.1) in \cite{Whitehead1992}. We also tested 
$\widetilde{\mu}_{CMLc}=\widehat{\mu}_{CML}-bias_{CML}(\widehat{\mu}_{CML}, R)$, but the results were almost identical  (data not shown).

\section{Comparison between conditional estimators}
\label{sec_comp}
\label{ch4}
When we compare the conditional estimators discussed above with the unconditional MLE $\widehat \mu_{ML}$, we can show the following theorem. The proof can be found in Web Appendix A.


\textbf{Theorem} Let $\widehat{\mu}\in \{\widehat{\mu}_{RB}, \widehat{\mu}_{CMU}, \widehat{\mu}_{CML}\}$.
If $R=1$, then 
$$
\widehat\mu \left\{ 
 \renewcommand{\arraystretch}{0.8}
  \begin{array}{l}
    < \\
    = \\
    > 
  \end{array} \right\} \widehat \mu_{ML}
  \mbox{ if } \widehat \mu_{ML} \left\{ 
 \renewcommand{\arraystretch}{0.8}
 \begin{array}{l}
    < \\
    = \\
    > 
  \end{array} \right\} \frac{c_1+c_2}{2}.
$$
If $R=2$ then $\widehat \mu < \widehat \mu_{ML}$.

\begin{figure}[htbp]
\centering
\includegraphics[width=160mm]{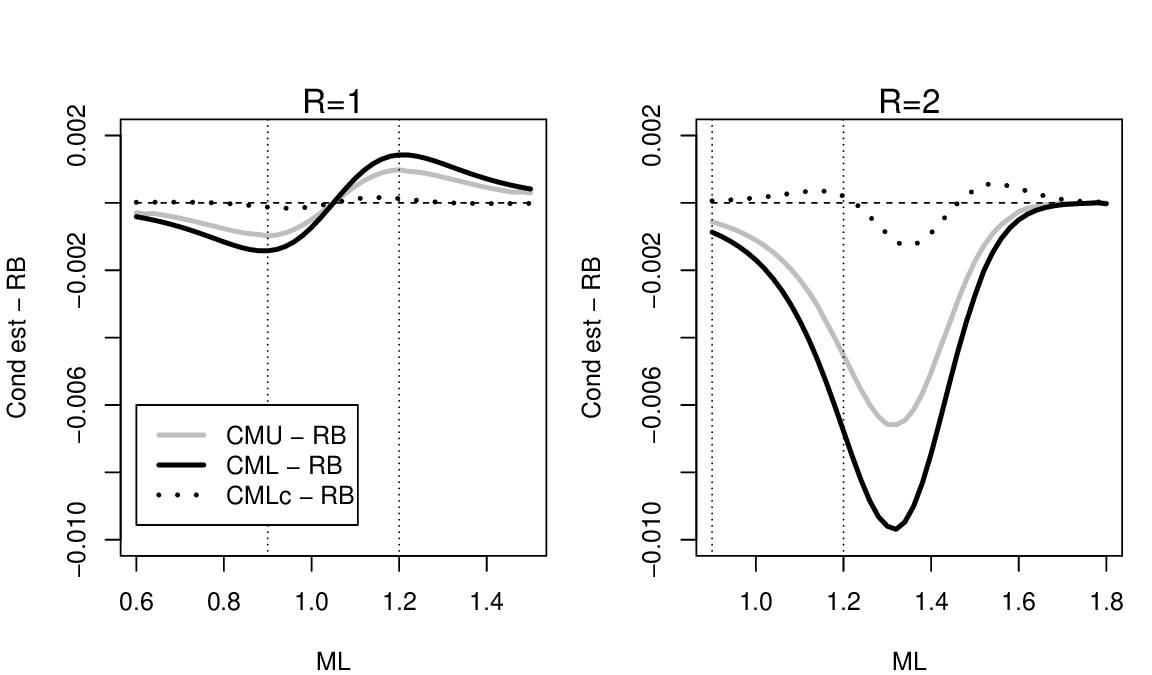}
\caption{Difference between conditional estimators and Rao-Blackwell (RB). Grey line:  conditional median unbiased - RB; Black line: conditional maximum likelihood (CML) - RB; Black dotted line: corrected CML - RB. All plotted versus maximum likelihood (ML) estimate. Here $n_{1}=50, n_{0}=100, n_{max}=150, c_{1}=0.9, c_{2}=1.2, \sigma=1$. The vertical dotted lines indicate $c_1$ and $c_2$.}
\label{fig_estimators}
\end{figure}

We calculate $\widehat \mu_{RB}$ as a function $g$ of $\widehat \mu_{ML}$, see (\ref{eq_rb}), and visualize the difference between $\widehat \mu_{RB}$ and $\widehat \mu_{ML}$ in Figure \ref{fig_rb_est}. 
Next we investigate the difference between the conditional estimators, $\widehat \mu_{RB}, \widehat \mu_{CMU}, \widehat \mu_{CML}, \widehat \mu_{CMLc}$.
Through formulae (\ref{cmu_def}), (\ref{eq_cml}), we have implicit representations of $\widehat \mu_{CMU}, \widehat \mu_{CML}$ as functions of $\widehat \mu_{ML}$, respectively; formula (\ref{eq:bias}) together with (\ref{eq_cml}) implicitly define  $\widehat \mu_{CMLc}$ as a function of $\widehat \mu_{ML}$. We have computed these functions numerically and show pairwise differences between the conditional estimators in Figure \ref{fig_estimators}, here for $n_{1}=50, n_{0}=100, n_{max}=150, c_{1}=0.9, c_{2}=1.2, \sigma=1$. The largest difference between $\widehat \mu_{CMU}$ and $\widehat \mu_{RB}$ in this scenario is for $R=2$ and $\widehat\mu_{ML}\approx 1.32$ and is $-0.0066$. Compared to the difference between $\widehat \mu_{RB}$ and $\widehat \mu_{ML}$ seen before, this is a quite small difference. The difference between $\widehat \mu_{CML}$ and $\widehat \mu_{RB}$ is a little larger but still at most 0.01. We see that we can order the estimators in terms of how much they correct the ML: $\widehat \mu_{RB}, \widehat \mu_{CMU}, \widehat \mu_{CML}$ (from smaller to larger absolute difference to MLE). This is parallel to the unconditional case where the median unbiased estimator is a compromise between the mean unbiased estimator and the unconditional MLE, see \cite{Bretz2009}, Section 6.2. The bias-corrected $\widehat \mu_{CMLc}$ is very similar to $\widehat \mu_{RB}$.

When analysing differences between estimators for other scenarios, we see that these are often very small. The estimators differ slightly more when $n_0-n_1$ and $n_{max}-n_0$ are smaller. In Web Appendix D, we show the differences for a few other scenarios.

\subsection{Simulation study}
\label{simstudy}
To assess average performance of the estimators, we simulated normal data for a number of scenarios and replicated a large number of times (1,000,000). In each replication the outcome of $Y_1$ defined $R$ and thereby decided the final design according to formula (\ref{three}). Without loss of generality 
$\sigma$ was kept fix at 1. We also fixed the value $c_1=0.9$ as results depend only on the differences between the values $\mu, c_1$ and $c_2$. We then varied $\mu$ over $\{0.9, 1, 1.2, 1.4\}$ and $c_{2}$ over $\{1.2,1.3\}$. The maximal patient number (used when $R=1$) was defined as $n_{max}=150$ throughout, and $n_0$ was set to $n_1+50$, while $n_1$ varied over  $\{50,70\}$. These scenarios are presented in Table \ref{tab1}, and some further scenarios in Web Appendix F and G.

In Table \ref{tab1} the methods are denoted as follows: Rao-Blackwell (RB), conditional median unbiased (CMU), conditional MLE (CML), bias corrected conditional MLE (CMLc), unconditional MLE (ML). 
In the simulations the CMLc used the correction $bias_{CML}(\widehat{\mu}|R)$ from equations (\ref{eq:biasdef}) and (\ref{eq:bias}). 
The table provides observed bias, variance and mean squared error (MSE) conditional on $R=1$ and $R=2$ as well as the number $s$ of replications with $R=1$ and $R=2$, respectively.

\begin{table}[p!]
\caption{Summary of bias and variation. Here $n_{0}=n_1+50, n_{max}=150, c_{1}=0.9, \sigma=1$. 1000,000 simulations. The values for $\mu$, $n_1$, $c_2$ are chosen in 4 scenarios according to the values noted in the table. The number of simulation runs with $R=1$ or $R=2$ is $s$.}\label{tab1}
\begin{center}
\begin{tabular}{cccllrrr}
  \hline
$\mu$ & $n_1$ & $c_2$ & Case & Method & Bias & Var & MSE \\ 
  \hline
1 & 50 & 1.2 & R=1 & RB & 0.000 & 0.009 & 0.009 \\ 
& & &   s= 682108 & CMU & -0.000 & 0.009 & 0.009 \\ 
& & &    & CML & -0.000 & 0.009 & 0.009 \\ 
& & &    & CMLc & 0.000 & 0.009 & 0.009 \\ 
& & &    & ML & 0.011 & 0.005 & 0.005 \\ 
& & &   R=2 & RB & 0.001 & 0.017 & 0.017 \\ 
& & &   s= 78778 & CMU & -0.002 & 0.017 & 0.017 \\ 
& & &    & CML & -0.004 & 0.017 & 0.017 \\ 
& & &    & CMLc & 0.001 & 0.017 & 0.017 \\ 
& & &    & ML & 0.132 & 0.006 & 0.023 \\  
   \hline
1.2 & 70 & 1.2 & R=1 & RB & -0.000 & 0.010 & 0.010 \\ 
& & &   s= 494010 & CMU & 0.002 & 0.010 & 0.010 \\ 
& & &    & CML & 0.003 & 0.010 & 0.010 \\ 
& & &    & CMLc & 0.000 & 0.010 & 0.010 \\  
& & &    & ML & -0.043 & 0.005 & 0.006 \\ 
& & &   R=2 & RB & 0.000 & 0.013 & 0.013 \\ 
& & &   s= 499857 & CMU & -0.007 & 0.013 & 0.013 \\ 
& & &    & CML & -0.010 & 0.013 & 0.013 \\ 
& & &    & CMLc & -0.000 & 0.013 & 0.013 \\ 
& & &    & ML & 0.056 & 0.005 & 0.008 \\ 
   \hline
1.4 & 50 & 1.3 & R=1 & RB & 0.000 & 0.009 & 0.009 \\ 
& & &  s= 239373 & CMU & 0.001 & 0.009 & 0.009 \\ 
& & &    & CML & 0.002 & 0.009 & 0.009 \\ 
& & &    & CMLc & 0.000 & 0.009 & 0.009 \\ 
& & &    & ML & -0.061 & 0.005 & 0.009 \\ 
& & &     R=2 & RB & 0.000 & 0.013 & 0.013 \\ 
& & &  s= 760435 & CMU & -0.005 & 0.013 & 0.013 \\ 
& & &    & CML & -0.008 & 0.013 & 0.013 \\ 
& & &    & CMLc & -0.001 & 0.013 & 0.013 \\ 
& & &    & ML & 0.029 & 0.008 & 0.009 \\ 
   \hline
0.9 & 50 & 1.2 & R=1 & RB & -0.000 & 0.009 & 0.009 \\ 
& & &  s= 482187 & CMU & -0.001 & 0.009 & 0.009 \\ 
& & &    & CML & -0.001 & 0.009 & 0.009 \\ 
& & &    & CMLc & -0.000 & 0.009 & 0.009 \\ 
& & &    & ML & 0.035 & 0.005 & 0.006 \\ 
& & &  R=2 & RB & -0.001 & 0.018 & 0.018 \\ 
& & &  s= 16968 & CMU & -0.003 & 0.018 & 0.018 \\ 
& & &    & CML & -0.005 & 0.017 & 0.017 \\ 
& & &    & CMLc & -0.001 & 0.018 & 0.018 \\ 
& & &    & ML & 0.175 & 0.005 & 0.036 \\  
   \hline
\end{tabular}
\end{center}
\end{table}

Among the methods there is a high degree of concordance. From a practical point of view one could argue that MSE should take precedence over both bias and variance. As judged by MSE the methods RB, CMU, CML, and CMLc perform equally well. If we require strict conditional unbiasedness, then of course Rao-Blackwell is the only option. 

However, we observe that the naive unconditional MLE does quite well. This has been previously mentioned in the case of sample size recalculation, see \cite{SIM:SIM2258}. In our simulations, the MLE performs best under $R=1$. If the true mean $\mu$ is large, 
the MLE also performs well under $R=2$ (see second and third scenario in Table \ref{tab1}). In cases when $\mu$ is small
but  $R=2$ happened anyway, the conditional MSE of the MLE is larger than for the other estimators (see first and last scenario in Table \ref{tab1}). Note that $R=2$ happened in $8\%$ and $2\%$ of the simulations in these two scenarios, respectively.

We have compared the simulated bias in Table \ref{tab1}  with the theoretical biases from (\ref{eq:biasml}), (\ref{eq:cmubias}) and (\ref{eq:biasdef}). The theoretical biases for ML (\ref{eq:biasml}) and CML (\ref{eq:biasdef}) are confirmed by the simulations. In contrast, approximation (\ref{eq:cmubias}) based on the delta method appears inaccurate.  

\subsection{Asymptotic distribution}
Noting that, as $n_{1} \rightarrow \infty$, $\mu \in (c_{1}, c_{2})$ implies $P(R=1) \rightarrow 1$, and, $\mu > c_{2}$ implies $P(R=2) \rightarrow 1$, one may prove the asymptotic conditional normality of $\widehat{\mu}_{ML}$ as $n_{1},n_{2}\rightarrow \infty$. These asymptotics show up in several of the simulation scenarios, see Web Appendix I. In some cases, slight deviations from the normal distribution are still visible for the finite sample sizes.
Regarding $\widehat \mu_{RB}=g_{i}(\widehat \mu_{ML})$ (for $R=i, i=1,2$), we note that $g_i$ is close to linear. The asymptotic conditional normality of $\widehat \mu_{ML}$  therefore leads to the asymptotic conditional normality of $\widehat \mu_{RB}$. It appears in the simulation results in Web Appendix I that asymptotic normality is attained, in some scenarios still with slight deviations for the sample sizes used.  

$\widehat{\mu}_{CMU}$ also roughly follows a conditional normal distribution in the simulation scenarios. We provide a heuristic argument for this. The distribution of $\widehat{\mu}_{ML}$ will asymptotically approach a normal distribution, which has $\mu$ as its median, thus $\widehat{\mu}_{CMU} \approx  \widehat{\mu}_{ML}$ for large samples.
Finally, it can be proven that $\sqrt{N}\mathscr{L}^{'}$ and consequently $\widehat{\mu}_{CML}$ tends to a conditional normal distribution as $n_{1}, n_{2} \rightarrow \infty$ \cite[Chapter 3, p.~86]{cox1994}.

We considered $n_{1},n_{2}\rightarrow \infty$ here. \cite{lane2012} argue for investigating asymptotics for finite $n_1$ and $n_{2}\rightarrow \infty$. They show that the unconditional distribution of the MLE in their case converges to a scale mixture of normal random variables. 

\section{Extensions}
\label{sec_extensions}
Next we discuss several possible generalizations of our basic model.

Consider first the case of unknown variance. Assuming a reasonable sample size, such as in our scenarios, the common variance is well estimated by the pooled variance estimate 
$S_{p}^{2}=\frac{(n_{1}-1)S_{1}^{2}+(n_{2}-1)S_{2}^{2}}{n_{1}+n_{2}-2}$, 
based on the stage-wise variance estimates $S_{1}^{2}$ and $S_{2}^{2}$. In a new set of simulations we inserted $S_p^2$ instead of $\sigma^2$ when computing the conditional estimators in the final analysis. For all 4 scenarios the values for Bias, Var, MSE were at most 0.001 different from the known-variance case (see Web Appendix E). For Scenario 4 with $R=2$ only, the differences were up to 0.004, which still can be explained by Monte-Carlo error as the number of repetitions under condition $R=2$ was smaller. 

The methods of Section \ref{ch3} extend in an obvious way to any one-parameter MLE $\widehat{\theta}(X)$, estimating an unknown parameter $\theta$, noting that  $P(R=1)=P\left\{c_1<\widehat{\theta}(X) < c_2\right\}$ approximately is $\Phi\left(-j(\widehat{\theta})(c_2-\widehat{\theta})\right)-\Phi\left(-j(\widehat{\theta})(c_1-\widehat{\theta})\right)$, cf. (\ref{phid}).

One may further extend our results concerning the one-arm model to two arms by the use of profile likelihood. In a parallel group study the estimation problem may be expressed as follows, cf. \cite{Whitehead1992}. The two independent series of observations obey laws $f(x, \mu_{1},\eta)$ and $f(x, \mu_{2},\eta)$, respectively, where $\eta$ is a vector of common nuisance parameters. For instance, $\sigma$ could be one of the nuisance parameters. The parameter of interest is the contrast $\theta=\frac{1}{2}(\mu_{1}-\mu_{2})$, while  $\psi=\frac{1}{2}(\mu_{1}+\mu_{2})$ becomes one additional nuisance parameter. 

Moreover, as in \cite{bebu2013} the scenario may be modified to accommodate adjustment for covariates. For simplicity  consider a single covariate $t$. Assume the individual outcomes $O_{i}=\bmath{x}_{i}\bmath{\beta}^{'}+\epsilon_{i}$ with $\bmath{\beta}=(\beta_{1}, \beta_{2},\beta_{3})$ and i.i.d. $\epsilon_{i} \sim N(0, \sigma^2)$, $\bmath{x}_{i}=(\delta_{i1},\delta_{i 2},t_{i})$, where $\delta_{ij}=1$ if subject $i$ receives treatment $j$ and $0$ otherwise. We may then replace $Y_{1}$ by $\widehat{\beta}^{(1)}_1-\widehat{\beta}^{(1)}_2$ with $\widehat{\bmath{\beta}}^{(1)}=(\widehat{\beta}^{(1)}_1, \widehat{\beta}^{(1)}_2,\widehat{\beta}^{(1)}_3) \sim N(\bmath{\beta}, (\mathbf{X}_{(1)}^{'}\mathbf{X}_{(1)})^{-1}\sigma^2))$, where $\mathbf{X}_{(1)}$ denotes the design matrix corresponding to the first stage, and $Y_{2}$ and $Y$ by the corresponding quantities.

Applying normal approximation (see \cite{Whitehead1992}, Chapter 3.4, or  \cite{jennisonturnbull}, Chapter 3.7), our results can be used in the situation of a two-arm time-to-event trial. Here, $n_1, n_0, n_{\max}$ are the number of events in  both arms rather than number of patients.
The normal approximation implies that $-\log(\widehat{HR}) \sim N(-\log(HR),4/n)$ where $HR$ is the true and $\widehat{HR}$ the estimated hazard ratio. Therefore, in our notation, we have $\sigma^2=4$. Note that we assume here additionally to the normal approximation that the ``increment'' from Stage 1 to Stage 2 is independent of Stage 1 in the test statistic. This assumption might be challenged in time-to-event trials (see \cite{jenkins2011}) as the same patient could e.g.\ contribute as a censored patient to Stage 1 and with an event to the increment. Hence, the independence assumption is an approximation.
We illustrate the application of our results to a time-to-event study in Section \ref{ch_ex}.

\section{Application in a schizophrenia trial}
\label{ch_ex}
In a randomized, double-blind trial for clinically stable patients with schizophrenia, the treatments quetiapine and placebo were compared, \cite{peuskens2007}. The primary endpoint was time to first schizophrenic relapse analysed by a Cox proportional hazards model. An interim analysis was conducted after 45 relapses observed in total
and the final analysis was scheduled after 90 relapses. (We ignore here for now the second interim 
which was preplanned -- we will discuss this below.) We apply normal approximation to the survival analysis data and the total number 
of relapses are treated as sample size.

The trial was preplanned to be stopped when the interim two-sided p-value was $<p_0=0.004455$, i.e.\ stop for efficacy if $p<p_0$ to the advantage of 
quetiapine, and stop for futility if $p<p_0$ to the advantage of placebo. The recruitment was fast, and  61 relapses therefore been 
observed when the independent Data and Safety Monitoring Board's recommendation ``stop'' or ``continue'' from the interim analysis 
could be announced. The number of 61 relapses was not preplanned; nevertheless, we treat the increase to 61 relapses as an unavoidable consequence of the 
recruitment process, as it had been planned in advance to have $n_0=61$ relapses. 

Using the approximation $-\log(\widehat{HR}) \sim N(-\log(HR),4/n)$, 
we have rule (\ref{three}) with $n_1=45, n_0=61, n_{\max}=90, \sigma^2=4$. We obtain
$
  c_1 = \frac{\sigma}{\sqrt{n_1}} \Phi^{-1}(p_0/2) = -0.848, c_2 = -c_1 = 0.848.
$

According to \cite{peuskens2007}, the observed hazard ratio was 0.16 after 45 relapses and 0.13 after 61 relapses.  We have $Y_1=-\log(\widehat{HR}_1)=1.83$ 
in the interim and $\widehat \mu_{ML}=2.04$ in the final analysis. In order to compute the Rao-Blackwell estimator, we compute 
first
$
  \sigma_1 = \sigma/\sqrt{n_1} = 2/\sqrt{45},  \sigma_2 = \sigma/\sqrt{N_2} = 1/2, 
$
and  
$
  \sigma_A^2 = \sigma_1^4/(\sigma_1^2+\sigma_2^2) = 0.0233, \sigma_B^2 = \sigma_2^4/(\sigma_1^2+\sigma_2^2) = 0.1844,
$
leading to 
$
  \widehat \mu_{RB}  
  = \widehat \mu_{ML} - 10^{-14} \approx \widehat \mu_{ML}.
$
Computing the other conditional estimates shows that in this case with a very clear result and $\widehat \mu_{ML}$ far from $c_2$, the estimates $\widehat \mu_{CMU}, \widehat \mu_{CML}, \widehat \mu_{CMLc}$ are almost equal to $\widehat \mu_{ML}$.

Assume that, in contrast to the true trial results, the interim result had just been over the stopping boundary with an estimated hazard ratio of 0.42 and that after 61 relapses the same estimate was obtained ($Y_1=\widehat \mu_{ML}=0.87$). This gives $\widehat \mu_{RB} = \widehat \mu_{ML} - 0.304 = 0.566$ corresponding to a Rao-Blackwell estimated hazard ratio of $\exp(-0.566)=0.57$. The estimated hazard ratios with CMU, CML and CMLc are 0.59, 0.60, 0.57, respectively.

In this clinical trial, the result was a stop after interim, $R=2$, and we condition on this outcome in the analysis. 
It does not matter for the conditional estimates what the design of the study would have been like if $R=1$ had occurred. The above calculated estimates are therefore also valid for the real situation where a second interim analysis was preplanned.

We made following simulations for this setting: 
Firstly, we simulated normal data according to the approximations made here including the idealisation of an independent increments assumption. 
Secondly, we simulated the patients' event times assuming exponential distribution and proportional hazards and assuming a realistic recruitment speed. At the calendar time when $n_1=45$ events were obtained, the hazard ratio was estimated using Cox regression and $Y_1$ was defined as $-\log(\widehat{HR}_1)$. Here, all patients were censored at calendar date for interim analysis when they had an event time afterwards. Based on $Y_1$, the decision for $R$ and $N$ is made.
The calendar time is determined when $N$ events are obtained and the hazard ratio was estimated as previously. Letting $Y=-\log(\widehat{HR})$, we defined the increment statistics $Y_2=(N\cdot Y - n_1 \cdot Y_1) / (N-n_1)$.

The simulation results and more details about simulation assumptions are included in Web Appendix G. The two ways to simulate produced similar results where the variances and MSEs in case of time-to-event simulation were slightly larger. The correlation between Stage 1 estimate and increment from Stage 1 to 2 was small, suggesting that the independent increments assumption can be seen as approximately valid in this situation. However, in the case of other event-time distributions this is not necessarily fulfilled.

\section{Discussion}
\label{ch5}
The naive MLE performs well in the simulations presented here. When comparing the unconditional MLE used in the conditional setting with the conditional estimators, the former had in many but not in all scenarios smaller MSE. However the unconditional MLE does not possess any optimality features in the conditional inference setting. Therefore it is worthwhile searching for alternatives that satisfy certain optimality criteria relevant to the conditional inference.
The difference between the four conditional estimators was quite small in the scenarios considered. This was also reflected in terms of their bias (they had little or no bias) and their variance, which was similar. It is therefore not of primary importance which of these is chosen. The conditional Rao-Blackwell estimator has the advantage that it is unbiased in construction and has an explicit representation making computation simpler. However, it has to be pointed out that the situation might be different for nonlinear models where differences between these types of estimators could be more marked. 

We considered a sample size rule with general boundaries $c_{1} < c_{2}$. One can choose $c_1$ and $c_2$ to achieve a certain conditional or predictive power (see e.g. \cite{jennisonturnbull}, ch.\ 10). In practice, these boundaries can only be a guide for the decision about Stage 2 made by a data monitoring committee, which also needs to consider safety and other aspects.


In practice, when there is particularly great uncertainty about the variance, it is reasonable to allow the sample size to depend on the interim variance estimate as well. I.e.\ in the design specified in (\ref{three}), the sample size in case $R=1$ would be $n_{\max}(S_1^2)$ where $S_1^2$ is an interim variance estimate. Conditional effect estimates for this design are a topic for future research.

Our results are directly applicable for sample size rules (\ref{two}) with a finite number of steps, e.g. constant sample sizes between three  cut-points $c_1, c_2, c_3$.
However, if an increasing number of cut-points are introduced, we will condition on more and more information from Stage 1. In the extreme case (for a continuous sample size recalculation formula) one would condition on the interim observation $Y_1$. This would imply that the UMVCUE will be  $\widehat{\mu}_{RB}=Y_2$ as mentioned in Section \ref{sec32}.  \cite{bs2006} discuss the fact that the information from Stage 1 is no longer used for the conditional estimate.

We point out that the described methods require prespecification of the sample size recalculation rule. \cite{graf2016} determine the maximum bias and MSE of the unconditional MLE when the interim decision rule is not prespecified. They do this by identifying the worst case sample size rule.  

A limitation of this elaboration is that we assumed known variance. In practice, variance is not known  and our methods would be used with estimated variance. We have investigated this for the simulation study, 
where we see similar results to known variance; if the sample sizes are large, the stagewise means are approximately normally distributed. However, in general, alternatives to our assumption of normally distributed data, $t$-distribution or other, might be considered in future research.  


\section{Supplementary Materials}

Web Appendices referenced in Sections \ref{ch3}, \ref{ch4}, \ref{sec_extensions} and \ref{ch_ex} are available with this paper at the Biometrics website on Wiley Online Library: {\tt http://onlinelibrary.wiley.com/doi/10.1111/biom.12642/full}.
\vspace*{-8pt}


\section*{Acknowledgements}
We thank the reviewers for comments which improved this article.

\bibliographystyle{apalike}
\bibliography{AdaptiveInf}  
\label{lastpage}
\end{document}